%% file: main.tex
\DocumentMetadata{}
\documentclass[sigconf,nonacm]{acmart}
\usepackage{amsmath}

\usepackage{multirow}
\usepackage{soul}
\usepackage{algorithm}
\usepackage{algpseudocode}
\algnewcommand\Not{\textbf{!}}
\algnewcommand{\algorithmicforeach}{\textbf{for each}}
\algdef{SE}[FOR]{ForEach}{EndForEach}[1]
  {\algorithmicforeach\ #1\ \algorithmicdo}% \ForEach{#1}
  {\algorithmicend\ \algorithmicforeach}% \EndForEach

\AtBeginDocument{%
  \providecommand\BibTeX{{%
    \normalfont B\kern-0.5em{\scshape i\kern-0.25em b}\kern-0.8em\TeX}}}

\copyrightyear{2024}
\acmYear{2024}
\acmConference[CAMS 2024]{2nd Workshop on Computer Architecture Modeling and Simulation}{November 2, 2024}{Austin, Texas, USA}
\acmBooktitle{2nd Workshop on Computer Architecture Modeling and Simulation (CAMS 2024), November 2, 2024, Austin, Texas, USA}
\begin{document}

%%
%% The "title" command has an optional parameter,
%% allowing the author to define a "short title" to be used in page headers.
%\title{Need for Parallel: Accelerating a modern GPU simulator}
\title{Parallelizing a modern GPU simulator}

%%
%% The "author" command and its associated commands are used to define
%% the authors and their affiliations.
%% Of note is the shared affiliation of the first two authors, and the
%% "authornote" and "authornotemark" commands
%% used to denote shared contribution to the research.
\author{Rodrigo Huerta}
\email{rodrigo.huerta.ganan@upc.edu}
\orcid{0000-0003-0052-7710}
\affiliation{%
  \institution{Universitat Politècnica de Catalunya}
  \city{Barcelona}
  \country{Spain}
}

\author{Antonio González}
\email{antonio@ac.upc.edu}
\orcid{0000-0002-0009-0996}
\affiliation{%
  \institution{Universitat Politècnica de Catalunya}
  \city{Barcelona}
  \country{Spain}
}

%%
%% By default, the full list of authors will be used in the page
%% headers. Often, this list is too long, and will overlap
%% other information printed in the page headers. This command allows
%% the author to define a more concise list
%% of authors' names for this purpose.
\renewcommand{\shortauthors}{Huerta and González}

% My commands
\newcommand{\summaryNumCores}{16}
\newcommand{\summaryAvgSpeedup}{5.8x}
\newcommand{\summaryMaxSpeedup}{14x}
\newcommand{\summaryMaxTimeSequential}{five days}
\newcommand{\summaryMaxTimeReductionOpenMp}{12 hours}

%%
%% The abstract is a short summary of the work to be presented in the
%% article.
\begin{abstract}
\par
Simulators are a primary tool in computer architecture research but are extremely computationally intensive. Simulating modern architectures with increased core counts and recent workloads can be challenging, even on modern hardware. This paper demonstrates that simulating some GPGPU workloads in a single-threaded state-of-the-art simulator such as Accel-sim can take more than \summaryMaxTimeSequential{}. In this paper we present a simple approach to parallelize this simulator with minimal code changes by using OpenMP. Moreover, our parallelization technique is deterministic, so the simulator provides the same results for single-threaded and multi-threaded simulations. Compared to previous works, we achieve a higher speed-up, and, more importantly, the parallel simulation does not incur any inaccuracies. When we run the simulator with \summaryNumCores{} threads, we achieve an average speed-up of \summaryAvgSpeedup{} and reach \summaryMaxSpeedup{} in some workloads. This allows researchers to simulate applications that take \summaryMaxTimeSequential{} in less than \summaryMaxTimeReductionOpenMp{}. By speeding up simulations, researchers can model larger systems, simulate bigger workloads, add more detail to the model, increase the efficiency of the hardware platform where the simulator is run, and obtain results sooner.
\end{abstract}

%%
%% The code below is generated by the tool at http://dl.acm.org/ccs.cfm.
%% Please copy and paste the code instead of the example below.
%%
\begin{CCSXML}
<ccs2012>
   <concept>
       <concept_id>10010520.10010521.10010528</concept_id>
       <concept_desc>Computer systems organization~Parallel architectures</concept_desc>
       <concept_significance>500</concept_significance>
       </concept>
   <concept>
       <concept_id>10010520.10010521.10010528.10010534</concept_id>
       <concept_desc>Computer systems organization~Single instruction, multiple data</concept_desc>
       <concept_significance>500</concept_significance>
       </concept>
   <concept>
       <concept_id>10010520.10010521.10010528.10010536</concept_id>
       <concept_desc>Computer systems organization~Multicore architectures</concept_desc>
       <concept_significance>500</concept_significance>
       </concept>
   <concept>
       <concept_id>10010147.10010341</concept_id>
       <concept_desc>Computing methodologies~Modeling and simulation</concept_desc>
       <concept_significance>500</concept_significance>
       </concept>
   <concept>
       <concept_id>10010147.10010169</concept_id>
       <concept_desc>Computing methodologies~Parallel computing methodologies</concept_desc>
       <concept_significance>500</concept_significance>
       </concept>
 </ccs2012>
\end{CCSXML}

\ccsdesc[500]{Computer systems organization~Parallel architectures}
\ccsdesc[500]{Computer systems organization~Single instruction, multiple data}
\ccsdesc[500]{Computer systems organization~Multicore architectures}
\ccsdesc[500]{Computing methodologies~Modeling and simulation}
\ccsdesc[500]{Computing methodologies~Parallel computing methodologies}
%%
%% Keywords. The author(s) should pick words that accurately describe
%% the work being presented. Separate the keywords with commas.
\keywords{GPU, GPGPU, microarchitecture, simulation, OpenMP, parallelization, GPGPU-Sim, Accel-sim}

%%
%% This command processes the author and affiliation and title
%% information and builds the first part of the formatted document.
\maketitle

\input{1.Introduction}
\input{2.RelatedWork}
\input{3.Work}
\input{4.Evaluation}
\input{5.Conclusion}

%%
%% The acknowledgments section is defined using the "acks" environment
%% (and NOT an unnumbered section). This ensures the proper
%% identification of the section in the article metadata, and the
%% consistent spelling of the heading.
\begin{acks}
This work has been supported by the CoCoUnit ERC Advanced Grant of the EU’s Horizon 2020 program (grant No 833057), the Spanish State Research Agency (MCIN/AEI) under grant PID2020-113172RB-I00, the Catalan Agency for University and Research (AGAUR) under grant 2021SGR00383, and the ICREA Academia program. We also thank Aurora Tomás for suggesting some changes to the paper.
\end{acks}

%%
%% The next two lines define the bibliography style to be used, and
%% the bibliography file.
\bibliographystyle{ACM-Reference-Format}
\bibliography{sample-base}

\end{document}

%% file: 1.Introduction.tex
\section{Introduction}

\par
Computer architects use simulators to design new microarchitectures, conduct research, and propose new designs or optimizations to existing ones. In GPGPUs, one of the most popular simulators is Accel-sim \cite{accelsim}, based on GPGPUSim \cite{gpgpusimOriginal}. The Accel-sim framework can simulate modern NVIDIA GPU architectures such as Volta \cite{voltaPaper}, Turing \cite{turingPaper}, Ampere \cite{amperePaper}, Ada \cite{adaPaper}, or Hopper\cite{hopperPaper}.

\par
\begin{figure}[ht]
  \centering
  \includegraphics[trim={1.3cm 1.3cm 1.4cm 0.3cm},clip,width=8cm]{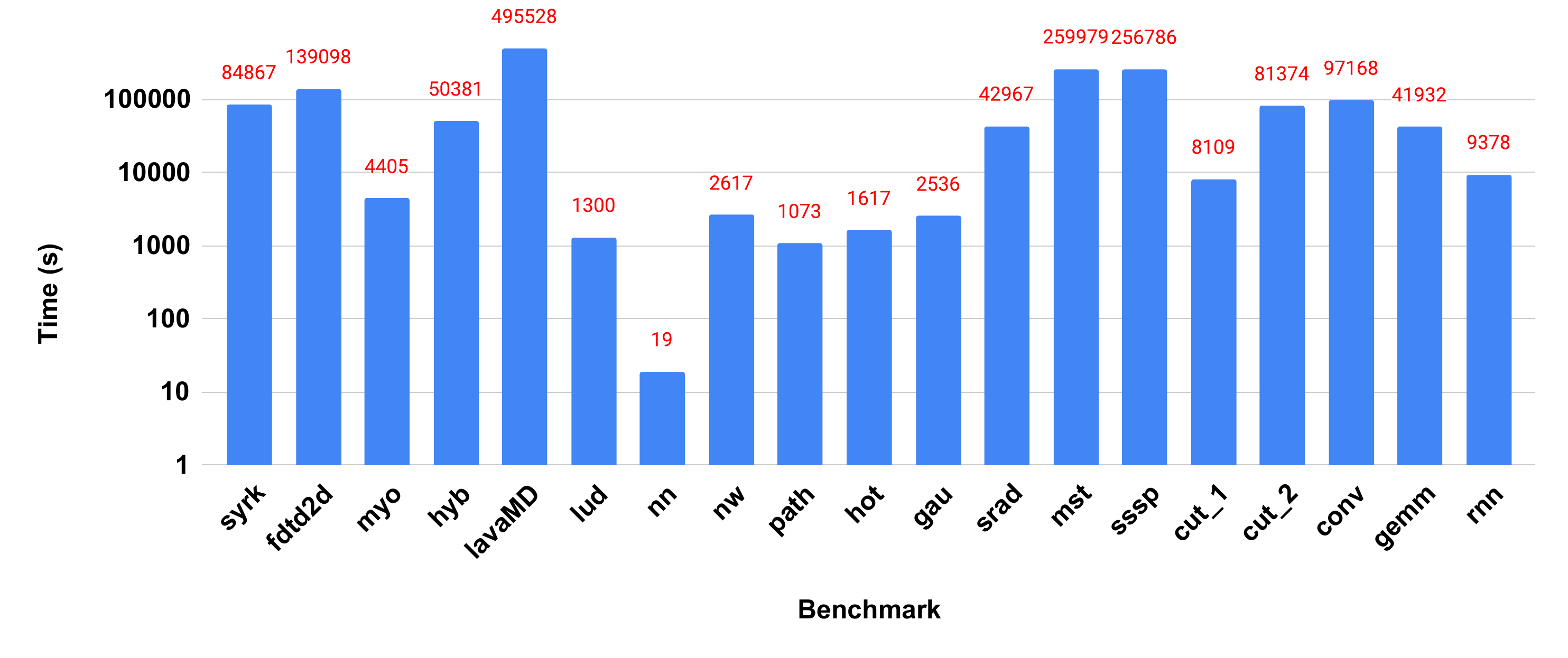}
  \caption{Time in seconds required to execute each workload with a single thread.}
  \label{fig:motivation_sim_time}
\end{figure}

Previous works have reported that simulating GPU systems can be 44,000x slower in Multi2Sim \cite{multi2sim} and 480,000x slower in GPGPU-Sim \cite{gputejas}. The Accel-sim framework is single-threaded, and simulating some workloads requires an immense amount of time to complete. \autoref{fig:motivation_sim_time} shows the time different applications take to be simulated in Accel-sim when executed in nodes with the specifications of Table 2. We can observe that many benchmarks require several hours to be simulated but some other applications such as \textit{mst}, \textit{sssp}, and \textit{lavaMD} need much more time, close to three days ($>259200s$) for \textit{mst} and \textit{sssp} and more than five days ($>432000s$) for \textit{lavaMD}. This long simulation runs compromise the amount of iterations in the typical research loop (propose a new feature, evaluate it) that can be performed and therefore, it limits the productivity of researchers.

\par
In this paper, we propose to parallelize the Accel-sim simulator with a simple approach using OpenMP \cite{openmp}. This achieves very important reductions in simulation time. For instance, using \summaryNumCores{} threads, we achieve an \summaryAvgSpeedup{} average speed-up and up to \summaryMaxSpeedup{} in some workloads. Besides, the user can easily configure how many threads each simulated workload will use, which provides several benefits. The main one is improving the efficiency of the hardware platform where simulations run by reducing the time unused cores are allocated by cluster schedulers such as SLURM \cite{slurm}, which assigns cores to jobs when they demand a certain amount of RAM even if they do not perform any computation. Moreover, our proposal does not affect simulation accuracy, a very critical aspect, unlike previous works that parallelize GPGPU-Sim \cite{parallelGPUSim1} \cite{parallelGPUSim2}. As the modifications to execute the simulator in parallel are minimal and can be easily configured to be disabled and executed sequentially, debug tasks are as easy as in the vanilla simulator. In addition, our approach is compatible with other techniques that speed up simulation by using sampling \cite{pcaKernelGpuSampling}, and others that reduce the detail of some components, such as NVAS \cite{nvas}.

\par
By speeding up simulators, researchers can model GPU architectures in more detail, simulate larger workloads and simulate bigger GPUs/multi-GPUs.

\par
The rest of this paper is organized as follows. In \autoref{sec:relatedwork}, we discuss previous works in the area. \hyperref[sec:work]{Section~\ref*{sec:work}} analyses the code of Accel-sim and explains how its parallelization has been implemented. Then, we evaluate the benefits of parallelizing the simulator in \autoref{sec:evaluation}. Finally, we summarize the main conclusions in \autoref{sec:conclusions}.

%% file: 2.RelatedWork.tex
\section{Related Work}\label{sec:relatedwork}

\par
Different GPU simulators have been developed to explore and propose architectural changes to these architectures. Some of the most popular ones are single-thread simulators, such as Multi2Sim \cite{multi2sim} or GPGPU-Sim \cite{gpgpusimOriginal}. The former models the AMD Evergreen \cite{amdevergreen} architecture, while the latter models the NVIDIA Tesla \cite{teslaHotchips}. Recently, GPGPU-Sim was updated and renamed as the Accel-sim framework \cite{accelsim} to include some major features introduced in the NVIDIA Volta \cite{voltaPaper} architecture.

\par
Some previous works have developed parallel GPU simulators. The first one is Barra \cite{barra}, a GPU functional simulator focused on the NVIDIA Tesla architecture, which achieves a speed-up of 3.53x with 4 threads. However, this simulator models an old architecture and does not provide a timing model. Another work that models the NVIDIA Tesla architecture is GpuTejas \cite{gputejas}, which includes a timing model and achieves a mean speed-up of 17.33x with 64 threads. Unfortunately, executing GpuTejas in parallel has an indeterministic behavior, leading to accuracy simulation errors of up to 7.7\% compared to the single-threaded execution. One of the most successful parallel simulators is MGPUSim \cite{mgpusim}, an event-driven simulator that includes functional and timing simulation targeting the AMD GCN3 \cite{amdgcn3}. MGPUSim follows a conservative parallel simulation approach for parallelizing the different concurrent events during the simulation, preventing any deviation error from executing the simulator in parallel. It achieves a mean speed-up of 2.5x when executed with 4 threads.

\par
Several works have parallelized the GPGPU-Sim simulator. MAFIA \cite{mafia} can run different kernels concurrently in multiple threads but cannot parallelize single-kernel simulations. Lee et al. \cite{parallelGPUSim1} \cite{parallelGPUSim2} have proposed a simulator framework built on top of GPGPU-Sim. Their proposal needs at least three threads in order to run. Two threads are always dedicated to executing the Interconnect-Memory Subsystem and the Work Distribution and Control components. The rest of the threads are devoted to parallelizing the execution of the multiple SMs of the GPU. Lee et al. approach has an average 3\% simulation error compared to the original sequential simulation, achieving an average speed-up of 5x and up to 8.9x in some benchmarks.

\par
Some simulators, such as NVAS \cite{nvas}, address the highly time-consuming problem of simulations by reducing the detail of some components. For example, modeling the GPU on-chip interconnects in low detail in NVAS reports a 2.13x speed-up and less than 1\% benefit in mean absolute error compared to a high-fidelity model. Avalos et al. \cite{pcaKernelGpuSampling} rely on sampling techniques to simulate huge workloads.

\par
In contrast to previous works, we follow a simple approach to parallelize the Accel-sim framework simulator, the most modern academic GPU simulator used for research and capable of executing modern NVIDIA GPU architectures and workloads. Our proposal employs OpenMP \cite{openmp} to implement a scalable implementation that allows parallelizing the simulator with a user-defined number of threads. Moreover, our approach does not compromise the simulation accuracy and determinism when the simulator runs in parallel and provides the same results as the sequential version. Thus, it eases developing and debugging tasks. This makes our work more robust than the implementations of Lee et al. \cite{parallelGPUSim1} \cite{parallelGPUSim2}, and GpuTejas \cite{gputejas}, where the parallel version results differ from the single-threaded one. Moreover, our work is orthogonal to approaches such as the ones followed by NVAS \cite{nvas} and Avalos et al. \cite{pcaKernelGpuSampling}, which reduce the detail of some components and use sampling to speed up simulations even more.

%% file: 3.Work.tex
\section{Parallelizing Accel-sim}\label{sec:work}

\par
This section describes how we have parallelized the Accel-sim framework simulator. 

\par
\begin{figure}[ht]
  \centering
  \includegraphics[trim={0.6cm 0.6cm 0.6cm 0.6cm},clip,width=8cm]{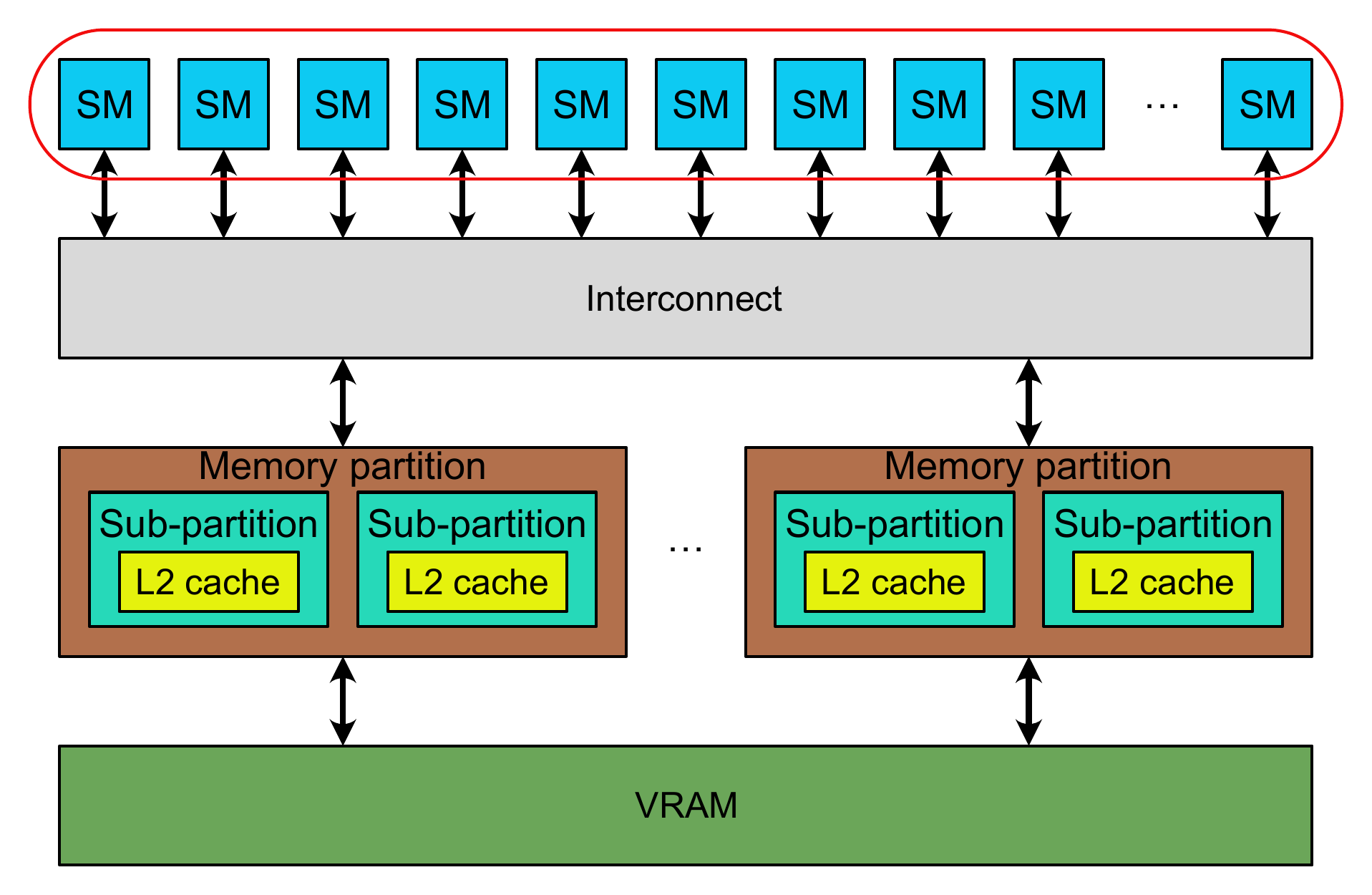}
  \caption{GPU design.}
  \label{fig:gpu_design_parallel}
\end{figure}

The principal components of the modeled GPUs in Accel-sim are shown in \autoref{fig:gpu_design_parallel}. The GPU has a dedicated main memory (VRAM), usually GDDR or HBM. There are several memory partitions, each with a channel to access the VRAM and the GPU's on-chip interconnect network. Memory partitions are divided into two sub-partitions, each with a slice of the L2 cache. Finally, there are a number of SMs (GPU cores) in charge of executing the user kernel instructions. 

\par
\begin{figure}[ht]
  \centering
  \includegraphics[trim={0.0cm 0.0cm 0.0cm 0.0cm},clip,width=6.5cm]{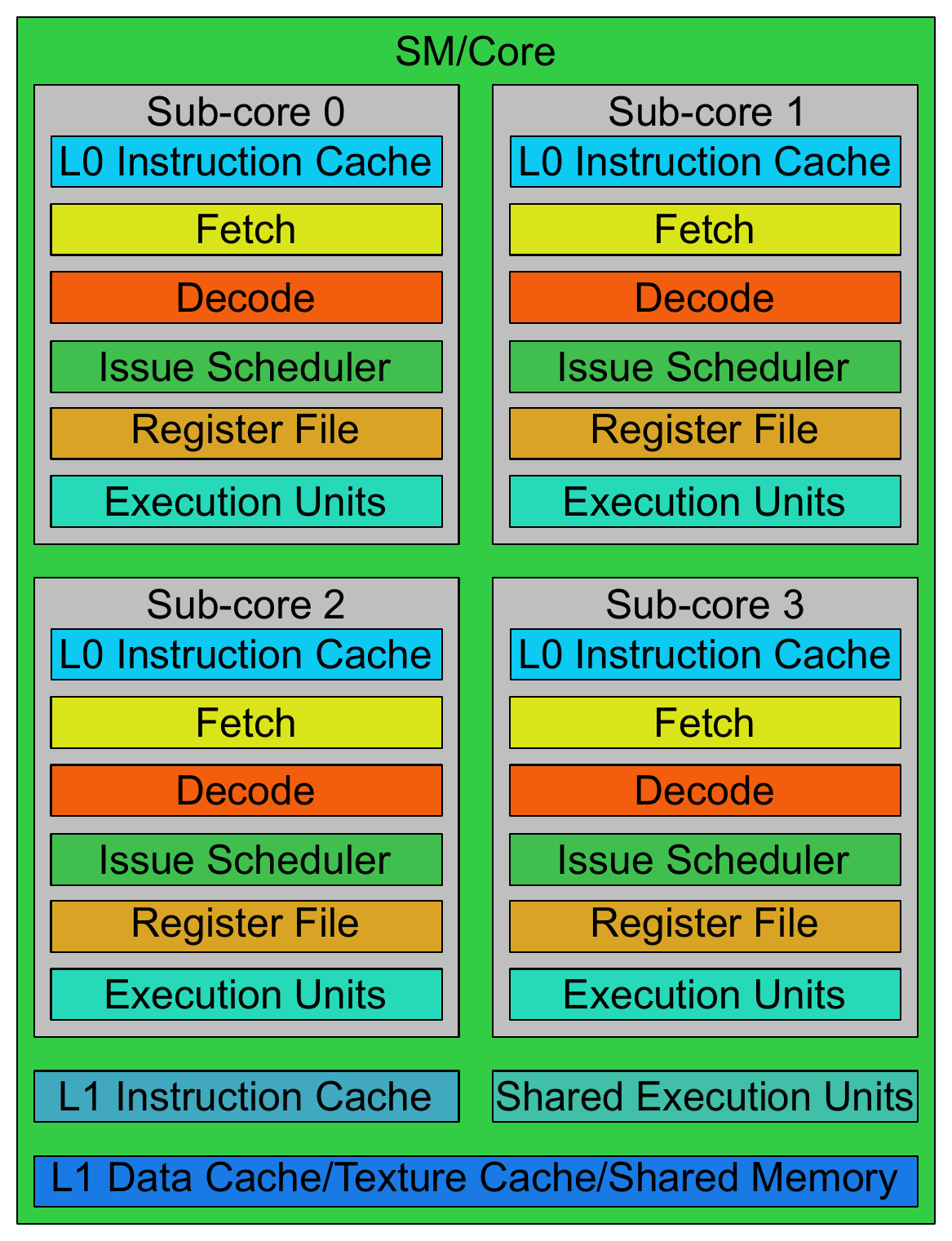}
  \caption{SM design.}
  \label{fig:gpu_design_sm}
\end{figure}

\autoref{fig:gpu_design_sm} shows the design of an SM. We can see that each core is divided into four sub-cores. They all have access to shared components such as an L1 instruction cache, the L1 Data Cache/Texture Cache/Shared memory, and shared execution units like FP64 for some architectures like Turing. Each sub-core is assigned a number of warps and executes them concurrently. The warp instruction fetcher requests one or several instructions from the L0 Instruction Cache every cycle. Once an instruction is received, it is decoded and stored in a buffer. An Issue Scheduler checks which warps have their oldest instruction ready and chooses one of them every cycle. Then, its operands are read from the register file, and finally, the instruction is usually executed in one of the different execution units of the sub-core (FP32, INT32, etc).

\par
\hyperref[alg:simulator]{Algorithm~\ref*{alg:simulator}} shows the high-level structure of the simulator's code to model the above-described architectures. We can see that the main function calls the cycle function while the simulation is still ongoing. The cycle function has different tasks to do. The first one processes all the interconnection network work; as we can see, this task is split into different code regions, lines 8-11, 16, and 19. It also models the main memory (lines 12-14) and the accesses to the L2 cache (line 16). Then, it continues by executing the work in each GPU's SMs (lines 21-23), each with the different components found in \autoref{fig:gpu_design_sm}. Finally, the function increments the number of cycles that the GPU is active and tries to issue the remainder thread blocks to available SMs.

\begin{algorithm}
\begin{algorithmic}[1]
\Function{main}{}
    \While{\Not simulation.done()}
        \State cycle()
    \EndWhile
\EndFunction
\State
\Function{Cycle}{}
    \State doIcntToSm()
    \ForEach{memSubpartition $\in$ GPU\_memSubpartition}
        \State doMemSubpartitionToIcnt()
    \EndForEach
    \ForEach{memPartition $\in$ GPU\_Partition}
        \State memPartition.DramCycle()
    \EndForEach
    \ForEach{memSubpartition $\in$ GPU\_memSubpartition}
        \State doIcntToMemSubpartition()
        \State memSubpartition.cacheCycle()
    \EndForEach
    \State doIcntScheduling()
    \State
    \ForEach{SM $\in$ GPU\_SMs}
        \State SM.cycle()
    \EndForEach
    \State gpuCycle++
    \State issueBlocksToSMs()
\EndFunction
\end{algorithmic}
\caption{Simulator pseudo-code}\label{alg:simulator}
\end{algorithm}

\begin{figure}[ht]
  \centering
  \includegraphics[trim={0.0cm 0.0cm 0.0cm 0.0cm},clip,width=8cm]{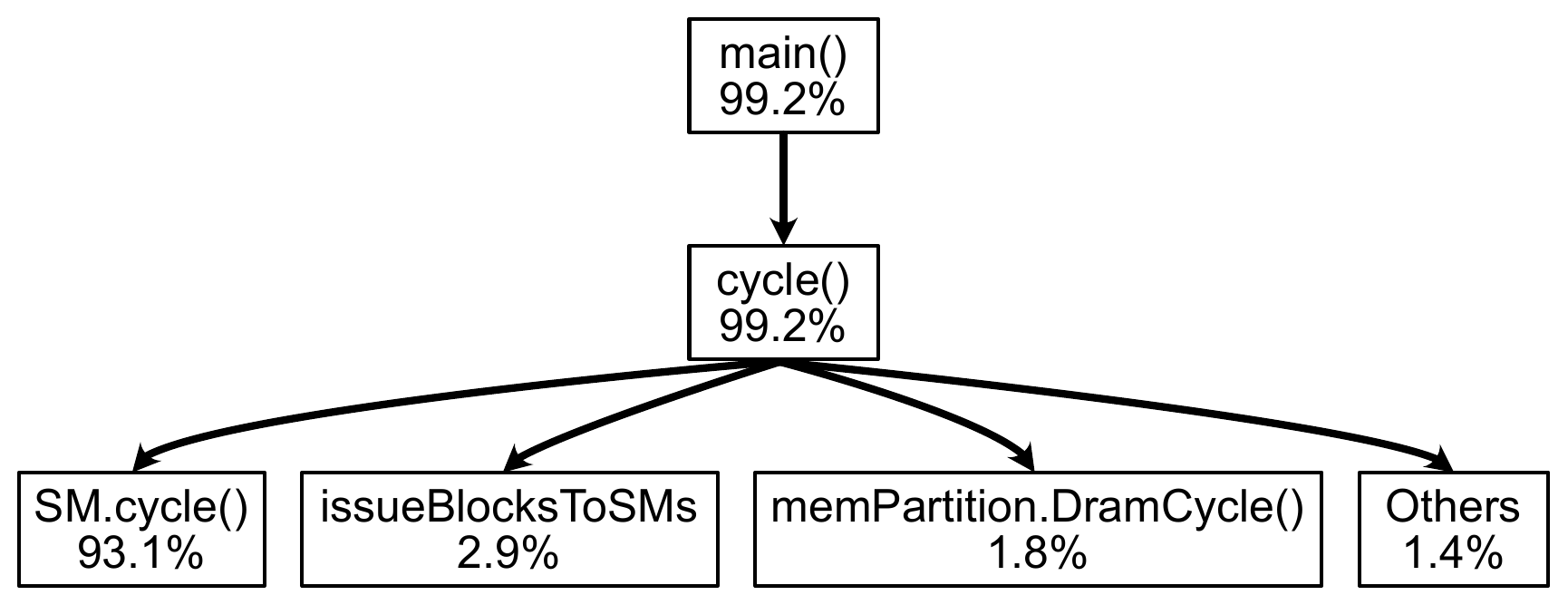}
  \caption{Profiler output.}
  \label{fig:gperftools}
\end{figure}

\par
To find out which parts of the simulator are the more time-consuming ones, we have configured the Google Performance Tools (Gperftools) \cite{gperftools} CPU profiler to be executed with the Accel-sim in a node with the specifications shown in \autoref{tab:node_specs}. The simulator models an NVIDIA RTX 3080 TI GPU (\autoref{tab:gpu_specs}) and simulates one of the benchmarks found in \autoref{tab:benchmarks} (concretely, \textit{hotspot}). \autoref{fig:gperftools} depicts the output of the Gperftools profiler, which shows that over $93\%$ of the execution time is spent executing the SM cycles. This makes sense due to two reasons. First, there are many more SMs than memory partitions. Second, an SM has many more components and details than memory partitions or DRAM. As a result, we have a clear target: parallelize the execution of all the SMs, which are circled in red in \autoref{fig:gpu_design_parallel}.

\par
We have parallelized the simulator using OpenMP because it requires minimal changes. First, we have added the \texttt{-fopenmp} flag to the simulator compilation. Then, we added the clause of OpenMP to parallelize for-loops in line 20
(\texttt{\#pragma omp parallel for}) of \hyperref[alg:simulator]{algorithm~\ref*{alg:simulator}}.

\par
Moreover, we had to fix the data races that appeared due to parallelizing the SMs loop. Although the different hardware components modeled in the SM were previously properly isolated,  stats had data races. Most of the stats of the Accel-sim simulator are shared among all the SMs to report a unique stat for the GPU. Usually, stats are counters that are later used to compute percentages or ratios. Therefore, we have isolated all these stats to be calculated by SM instead of globally for the whole GPU. Once the kernel execution has finished, each of the stats reported by SM is gathered into a single GPU stat to report stats in the same way as the single-threaded simulator. Notice that this approach is much better than creating a critical section whenever we want to increase a stat counter because this kind of construct would damage performance due to frequent code serialization and lock management \cite{criticalSectionOpenMP}.

\par
Even though counters are the most common stat, sometimes stats are represented by hash tables or sets. For example, suppose we want to discover how many different memory addresses are accessed during simulation. In that case, we need a set (which does not contain duplicates) that tracks all the accessed addresses. However, maps or sets are not thread-safe structures in C++ \cite{stlThreadSafe}, meaning they undermine behavior and can lead to segmentation fault errors. Therefore, there are three possible solutions to this problem. The first one is to make this structure thread-safe by ourselves. The second one is to have one of these structures per SM and then compute the union of all SM data structures at the end. The third one is to find a place where the simulator is executed sequentially and handle that stat there (e.g. making the different insert/erase operations). It is clear that the last option is the best one. However, it may not always be possible, and the simulator user will have to choose between the first or the second option in those scenarios. The choice will depend on a trade-off between the performance drop of accessing a unique shared thread-safe structure shared by all SMs by employing critical sections or increasing memory consumption by having per-SM data structures.

%% file: 4.Evaluation.tex
\section{Evaluation}\label{sec:evaluation}

\par
This section evaluates the performance benefits of parallelizing the Accel-sim framework simulator \cite{accelsim}. First, we describe our evaluation methodology to measure the speed-up of the parallel simulator. Then, we present a sensitive analysis of how the speed-up changes depending on the number of threads devoted to the execution. Finally, we analyze the impact of the for-loop OpenMP scheduler in the simulator.

\subsection{Methodology}

\par
We have configured the simulator with the parameters shown in \autoref{tab:gpu_specs}, which represent an NVIDIA RTX 3080 TI GPU based on the Ampere architecture. 

\begin{table}
    \centering
    %\tiny
    \begin{tabular}{ |c|c|  }
    \hline
shs 
Parameter & Value \\
    \hline
    Core Clock & 1365 MHz \\
    Mem. Clock & 9500 MHz \\
    \# SM & 80 \\
    \# Warps per SM & 48  \\
    Total Shared memory/L1D per SM & 128 KB \\
    \# Mem. part. & 24 \\
    Total L2 cache  & 6 MB \\
    \hline
    \end{tabular}
    \caption{NVIDIA RTX 3080 TI GPU simulator parameters.}
    \label{tab:gpu_specs}
\end{table}

\par
\autoref{tab:benchmarks} lists the different benchmark suites that we have employed to measure the efficacy of the parallelization. They represent a variety of very commonly used benchmarks that exhibit different degrees of parallelism.

\begin{table}
    \centering
    %\tiny
    \begin{tabular}{ |c|  }
    \hline
    \textbf{Rodinia 3.1} \cite{rodinia} \\
    \hline
    \multirow{2}{=}{gaussian (gau), hotspot (hot), hybridsort (hyb), lavaMD, lud, myocyte (myo), nn, nw, pathfinder (path), srad\_v1 (srad)} \\
    \\
    \hline
    \textbf{Polybench} \cite{polybench} \\
    \hline
    \multirow{1}{=}{fdtd2d, syrk} \\
    \hline
    \textbf{Lonestar} \cite{lonestar} \\
    \hline
    \multirow{1}{=}{mst, sssp} \\
    \hline
    \textbf{Deepbench} \cite{deepbenchWeb} \\
    \hline
    \multirow{1}{=}{conv, gemm, rnn} \\
    \hline
    \textbf{Cutlass} \cite{cutlass} \\
    \hline
    \multirow{1}{=}{2560x16x2560 (cut\_1), 2560x1024x2560 (cut\_2)} \\
    \hline
    \end{tabular}
    \caption{Benchmarks.}
    \label{tab:benchmarks}
\end{table}

\par
All the simulations have been executed in a cluster of homogeneous nodes with the specifications shown in \autoref{tab:node_specs}.

\begin{table}
    \centering
    %\tiny
    \begin{tabular}{ |c|c|  }
    \hline
    \multicolumn{2}{|c|}{CPU} \\
    \hline
    Model & AMD Epyc 7401P \\
    Cores & 24 \\
    Threads & 48 \\
    Frequency & 2 GHz \\
    \hline
    \multicolumn{2}{|c|}{RAM} \\
    \hline
    Total size & 128 GB \\
    Technology & DDR4 \\
    \hline
    \end{tabular}
    \caption{Node specifications.}
    \label{tab:node_specs}
\end{table}

\subsection{Parallel Speed-Up}

\par
This subsection analyzes how the speed-up evolves depending on the number of threads used by the simulator. 

\par
\begin{figure*}[ht]
  \centering
  \includegraphics[trim={1.4cm 0.6cm 0.6cm 0.6cm},clip,width=17.5cm]{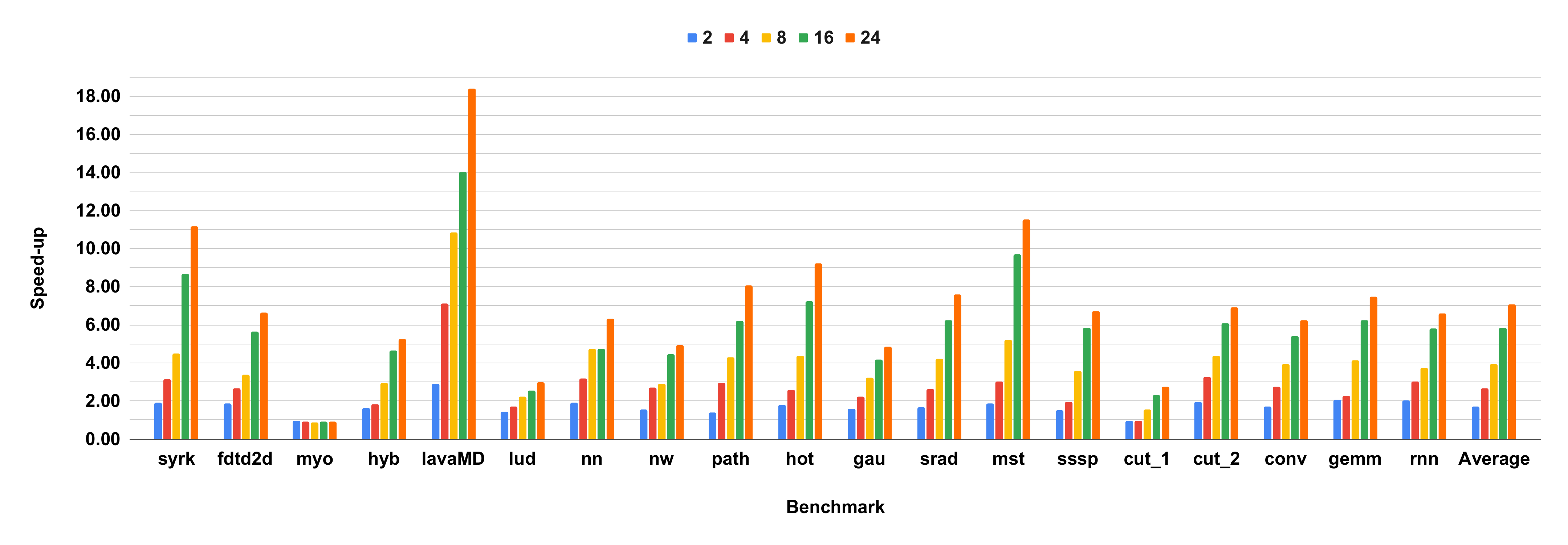}
  \caption{Speed-up with a different number of threads against the single-threaded version.}
  \label{fig:eval_sensitivity}
\end{figure*}

\autoref{fig:eval_sensitivity} shows the speed-up obtained with 2, 4, 8, 16, and 24 threads, averaging 1.72x, 2.64x, 3.95x, 5.83x, and 7.08x, respectively, against the single-threaded simulator. Executing the simulator with more than eight threads is less efficient: the efficiency is 0.36 for 16 threads, and 0.3 for 24 threads. However, some specific benchmarks, such as \textit{lavaMD}, significantly benefit from this high number of threads, reaching a speed-up of \summaryMaxSpeedup{} and an efficiency of 0.88 with 16 threads. This speed-up reduces the simulation slowdown compared to real hardware from 10,748,031x of the single-threaded simulator to 766,423x of the parallel one. Moreover, this benchmark is one of the most benefited from parallelization as it achieves super speed-up with 2, 4, and 8 threads configurations.

\begin{figure*}[ht]
  \centering
  \includegraphics[trim={0.6cm 0.6cm 0.6cm 0.6cm},clip,width=17.5cm]{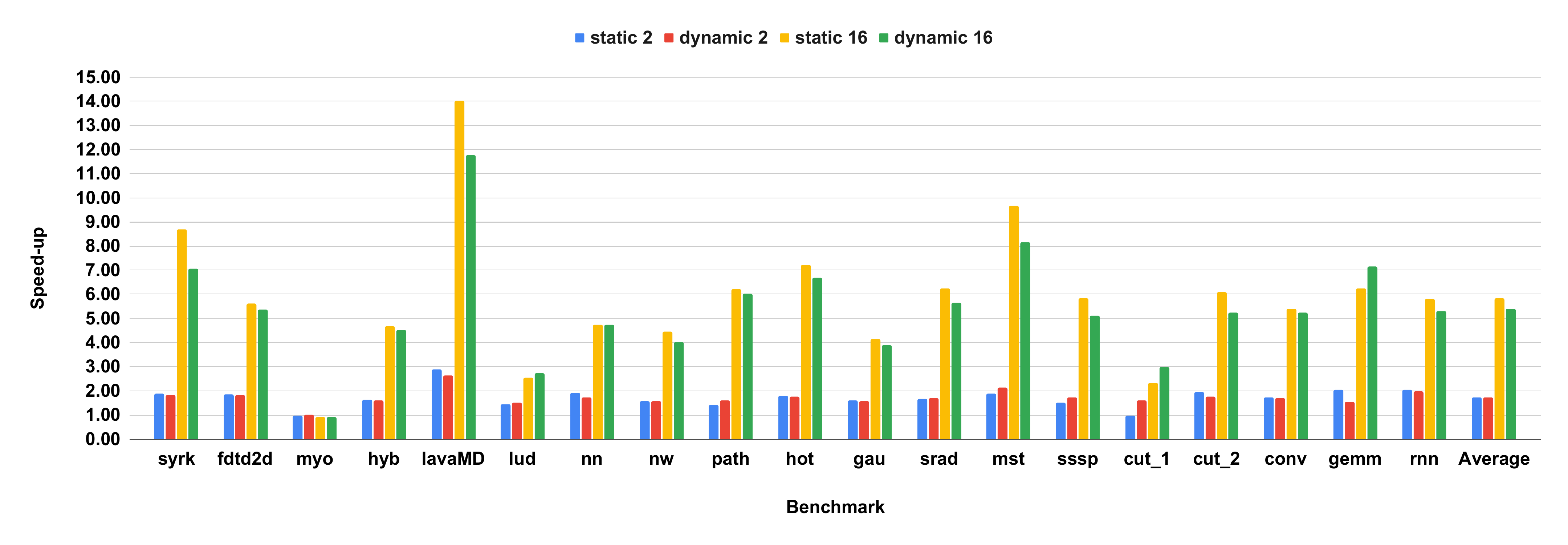}
  \caption{Speed-up obtained with the dynamic and static OpenMP for-loop scheduler with 2 and 16 threads against the single-threaded version.}
  \label{fig:eval_openmp_scheduler}
\end{figure*}

\begin{figure}[ht]
  \centering
  \includegraphics[trim={0.6cm 0.6cm 0.6cm 0.6cm},clip,width=8cm]{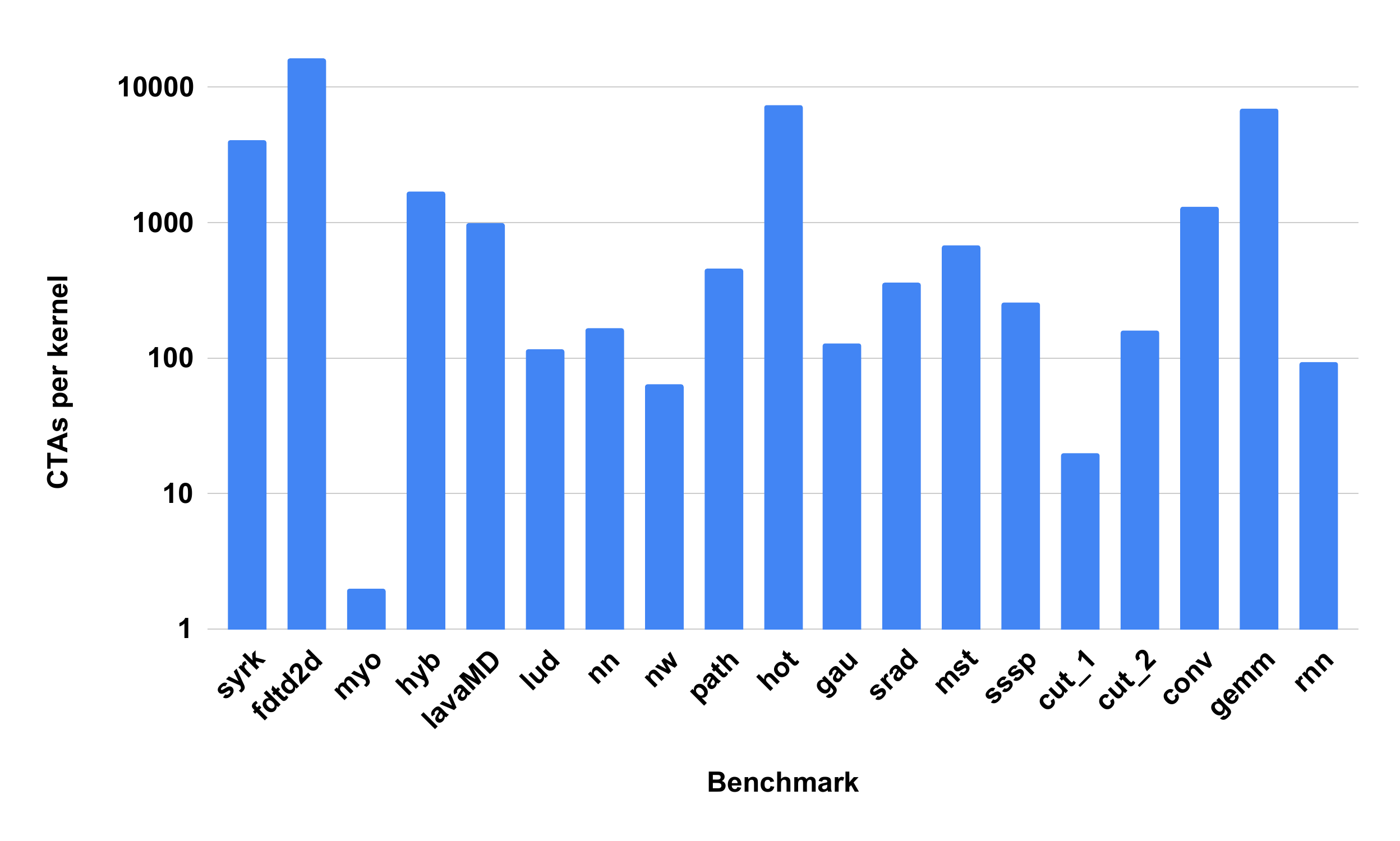}
  \caption{Number of CTAs per kernel.}
  \label{fig:ctas_per_kernel}
\end{figure}

\par
Other workloads, such as \textit{myocyte}, which has a tiny number of CTAs (thread blocks) per kernel (2), do not benefit from parallelization, and it is penalized by running the OpenMP interface, resulting in minor slowdowns. To understand why, we need to know that CTAs are distributed in a round-robin fashion among the GPU SMs. As there are only two CTAs, only two SMs are active during the simulation. Therefore, parallelizing the execution of the rest of the SMs is useless. As shown in \autoref{fig:ctas_per_kernel}, workloads usually have many more CTAs per kernel than \textit{myocyte}, and more than the GPU's number of SMs (80).

\par
Computing the correlation factor between the speed-up obtained with 16 threads and the time to execute a workload in a single-thread, reveals a strong positive correlation with a value of 0.78. This means that the more time the application needs to be simulated in a single-thread, the more benefit it gets from parallelizing the simulator.

\subsection{OpenMP scheduler}

\par
Previous works \cite{ompScheduling1} \cite{ompScheduling2} analyzed the impact of the OpenMP for-loop scheduler. There are two main OpenMP schedulers: static and dynamic. In a static scheduler, the iterations of a for loop are distributed statically to threads. Therefore, it has little overhead and fits perfectly in regular and balanced applications. On the other hand, the dynamic scheduler assigns work to the threads when they are idle. Thus, it fits better in unbalanced environments. However, the dynamic scheduler has bigger overheads than the static one because it distributes the iterations of the loop at runtime, so it performs worse in balanced environments.

\par
\autoref{fig:eval_openmp_scheduler} shows how the OpenMP scheduler affects the benefits of parallelization depending on the number of threads in use. Both schedulers are configured with a granularity of one in the iteration distribution.

\par
We can see that applications with a negligible number of CTAs per kernel, such as \textit{myocyte}, perform similarly and do not benefit from parallelism. However, other workloads with a small number of CTAs per kernel, such as \textit{cut\_1}, benefit a lot from having the dynamic scheduler. Concretely, \textit{cut\_1} goes from the 0.97x speedup with the static scheduler to 1.61x when running with two threads. Other regular and balanced benchmarks, such as \textit{cut\_2} or \textit{lavaMD}, consistently perform better with the static than with the dynamic, as it does not have the scheduler overheads. Finally, there are workloads such as \textit{sssp} that, depending on the number of threads, perform better with one scheduler or the other. 

%This is because increasing the number of threads decreases the criticality of load unbalance.

%% file: 5.Conclusion.tex
\section{Conclusions} \label{sec:conclusions}

\par
In this paper, we propose a simple technique for parallelizing the Accel-sim framework simulator \cite{accelsim}, which is one of the most popular tools for researching GPGPU architectures. We rely on the OpenMP \cite{openmp} application programming interface based on the shared memory paradigm. OpenMP allows us to parallelize the simulator with minimal changes. Our approach is deterministic and provides the same results when running the simulator single-threaded or multi-threaded. In other words, it does not incur simulation inaccuracies due to parallelization as some previous works do. When \summaryNumCores{} threads are in use, we achieve an average speed-up of \summaryAvgSpeedup{} and reach up to \summaryMaxSpeedup{} in some workloads. Our approach allows researchers that use Accel-sim to have several benefits, such as modeling bigger systems, simulating larger workloads, adding more detail to the GPU model, increasing simulation hardware efficiency, and obtaining results sooner.

%% file: main.bbl
%%% -*-BibTeX-*-
%%% Do NOT edit. File created by BibTeX with style
%%% ACM-Reference-Format-Journals [18-Jan-2012].

\begin{thebibliography}{31}

%%% ====================================================================
%%% NOTE TO THE USER: you can override these defaults by providing
%%% customized versions of any of these macros before the \bibliography
%%% command.  Each of them MUST provide its own final punctuation,
%%% except for \shownote{}, \showDOI{}, and \showURL{}.  The latter two
%%% do not use final punctuation, in order to avoid confusing it with
%%% the Web address.
%%%
%%% To suppress output of a particular field, define its macro to expand
%%% to an empty string, or better, \unskip, like this:
%%%
%%% \newcommand{\showDOI}[1]{\unskip}   % LaTeX syntax
%%%
%%% \def \showDOI #1{\unskip}           % plain TeX syntax
%%%
%%% ====================================================================

\ifx \showCODEN    \undefined \def \showCODEN     #1{\unskip}     \fi
\ifx \showDOI      \undefined \def \showDOI       #1{#1}\fi
\ifx \showISBNx    \undefined \def \showISBNx     #1{\unskip}     \fi
\ifx \showISBNxiii \undefined \def \showISBNxiii  #1{\unskip}     \fi
\ifx \showISSN     \undefined \def \showISSN      #1{\unskip}     \fi
\ifx \showLCCN     \undefined \def \showLCCN      #1{\unskip}     \fi
\ifx \shownote     \undefined \def \shownote      #1{#1}          \fi
\ifx \showarticletitle \undefined \def \showarticletitle #1{#1}   \fi
\ifx \showURL      \undefined \def \showURL       {\relax}        \fi
% The following commands are used for tagged output and should be
% invisible to TeX
\providecommand\bibfield[2]{#2}
\providecommand\bibinfo[2]{#2}
\providecommand\natexlab[1]{#1}
\providecommand\showeprint[2][]{arXiv:#2}

\bibitem[AMD(2011)]%
        {amdevergreen}
\bibfield{author}{\bibinfo{person}{AMD}.} \bibinfo{year}{2011}\natexlab{}.
\newblock \bibinfo{title}{AMD Evergreen Family Instruction Set Arch}.
\newblock
\newblock
\urldef\tempurl%
\url{www.amd.com}
\showURL{%
\tempurl}


\bibitem[AMD(2016)]%
        {amdgcn3}
\bibfield{author}{\bibinfo{person}{AMD}.} \bibinfo{year}{2016}\natexlab{}.
\newblock \bibinfo{title}{Graphics Core Next Architecture, Generation 3, Reference Guide}.
\newblock
\newblock
\urldef\tempurl%
\url{www.amd.com}
\showURL{%
\tempurl}


\bibitem[Avalos~Baddouh et~al\mbox{.}(2021)]%
        {pcaKernelGpuSampling}
\bibfield{author}{\bibinfo{person}{Cesar Avalos~Baddouh}, \bibinfo{person}{Mahmoud Khairy}, \bibinfo{person}{Roland~N. Green}, \bibinfo{person}{Mathias Payer}, {and} \bibinfo{person}{Timothy~G. Rogers}.} \bibinfo{year}{2021}\natexlab{}.
\newblock \showarticletitle{Principal Kernel Analysis: A Tractable Methodology to Simulate Scaled GPU Workloads}. In \bibinfo{booktitle}{\emph{MICRO-54: 54th Annual IEEE/ACM International Symposium on Microarchitecture}} (Virtual Event, Greece) \emph{(\bibinfo{series}{MICRO '21})}. \bibinfo{publisher}{Association for Computing Machinery}, \bibinfo{address}{New York, NY, USA}, \bibinfo{pages}{724–737}.
\newblock
\showISBNx{9781450385572}
\urldef\tempurl%
\url{https://doi.org/10.1145/3466752.3480100}
\showDOI{\tempurl}


\bibitem[Ayguad{\'e} et~al\mbox{.}(2003)]%
        {ompScheduling1}
\bibfield{author}{\bibinfo{person}{Eduard Ayguad{\'e}}, \bibinfo{person}{Bob Blainey}, \bibinfo{person}{Alejandro Duran}, \bibinfo{person}{Jes{\'u}s Labarta}, \bibinfo{person}{Francisco Mart{\'i}nez}, \bibinfo{person}{Xavier Martorell}, {and} \bibinfo{person}{Ra{\'u}l Silvera}.} \bibinfo{year}{2003}\natexlab{}.
\newblock \showarticletitle{Is the Schedule Clause Really Necessary in OpenMP?}. In \bibinfo{booktitle}{\emph{OpenMP Shared Memory Parallel Programming}}, \bibfield{editor}{\bibinfo{person}{Michael~J. Voss}} (Ed.). \bibinfo{publisher}{Springer Berlin Heidelberg}, \bibinfo{address}{Berlin, Heidelberg}, \bibinfo{pages}{147--159}.
\newblock
\showISBNx{978-3-540-45009-2}


\bibitem[Bakhoda et~al\mbox{.}(2009)]%
        {gpgpusimOriginal}
\bibfield{author}{\bibinfo{person}{Ali Bakhoda}, \bibinfo{person}{George~L. Yuan}, \bibinfo{person}{Wilson W.~L. Fung}, \bibinfo{person}{Henry Wong}, {and} \bibinfo{person}{Tor~M. Aamodt}.} \bibinfo{year}{2009}\natexlab{}.
\newblock \showarticletitle{Analyzing CUDA workloads using a detailed GPU simulator}. In \bibinfo{booktitle}{\emph{2009 IEEE International Symposium on Performance Analysis of Systems and Software}}. \bibinfo{pages}{163--174}.
\newblock
\urldef\tempurl%
\url{https://doi.org/10.1109/ISPASS.2009.4919648}
\showDOI{\tempurl}


\bibitem[Burtscher et~al\mbox{.}(2012)]%
        {lonestar}
\bibfield{author}{\bibinfo{person}{Martin Burtscher}, \bibinfo{person}{Rupesh Nasre}, {and} \bibinfo{person}{Keshav Pingali}.} \bibinfo{year}{2012}\natexlab{}.
\newblock \showarticletitle{{A quantitative study of irregular programs on GPUs}}. In \bibinfo{booktitle}{\emph{2012 IEEE International Symposium on Workload Characterization (IISWC)}}. \bibinfo{pages}{141--151}.
\newblock
\urldef\tempurl%
\url{https://doi.org/10.1109/IISWC.2012.6402918}
\showDOI{\tempurl}


\bibitem[Che et~al\mbox{.}(2009)]%
        {rodinia}
\bibfield{author}{\bibinfo{person}{Shuai Che}, \bibinfo{person}{Michael Boyer}, \bibinfo{person}{Jiayuan Meng}, \bibinfo{person}{David Tarjan}, \bibinfo{person}{Jeremy~W. Sheaffer}, \bibinfo{person}{Sang~Ha Lee}, {and} \bibinfo{person}{Kevin Skadron}.} \bibinfo{year}{2009}\natexlab{}.
\newblock \showarticletitle{{Rodinia: A benchmark suite for heterogeneous computing}}. In \bibinfo{booktitle}{\emph{Proceedings of the 2009 IEEE International Symposium on Workload Characterization, IISWC 2009}}. \bibinfo{pages}{44--54}.
\newblock
\showISBNx{9781424451562}
\urldef\tempurl%
\url{https://doi.org/10.1109/IISWC.2009.5306797}
\showDOI{\tempurl}


\bibitem[Ciorba et~al\mbox{.}(2018)]%
        {ompScheduling2}
\bibfield{author}{\bibinfo{person}{Florina~M. Ciorba}, \bibinfo{person}{Christian Iwainsky}, {and} \bibinfo{person}{Patrick Buder}.} \bibinfo{year}{2018}\natexlab{}.
\newblock \showarticletitle{OpenMP Loop Scheduling Revisited: Making a Case for More Schedules}. In \bibinfo{booktitle}{\emph{Evolving OpenMP for Evolving Architectures}}, \bibfield{editor}{\bibinfo{person}{Bronis~R. de~Supinski}, \bibinfo{person}{Pedro Valero-Lara}, \bibinfo{person}{Xavier Martorell}, \bibinfo{person}{Sergi Mateo~Bellido}, {and} \bibinfo{person}{Jesus Labarta}} (Eds.). \bibinfo{publisher}{Springer International Publishing}, \bibinfo{address}{Cham}, \bibinfo{pages}{21--36}.
\newblock
\showISBNx{978-3-319-98521-3}


\bibitem[Collange et~al\mbox{.}(2010)]%
        {barra}
\bibfield{author}{\bibinfo{person}{Caroline Collange}, \bibinfo{person}{Marc Daumas}, \bibinfo{person}{David Defour}, {and} \bibinfo{person}{David Parello}.} \bibinfo{year}{2010}\natexlab{}.
\newblock \showarticletitle{Barra: A Parallel Functional Simulator for GPGPU}. In \bibinfo{booktitle}{\emph{2010 IEEE International Symposium on Modeling, Analysis and Simulation of Computer and Telecommunication Systems}}. \bibinfo{pages}{351--360}.
\newblock
\urldef\tempurl%
\url{https://doi.org/10.1109/MASCOTS.2010.43}
\showDOI{\tempurl}


\bibitem[Dagum and Menon(1998)]%
        {openmp}
\bibfield{author}{\bibinfo{person}{L. Dagum} {and} \bibinfo{person}{R. Menon}.} \bibinfo{year}{1998}\natexlab{}.
\newblock \showarticletitle{OpenMP: an industry standard API for shared-memory programming}.
\newblock \bibinfo{journal}{\emph{IEEE Computational Science and Engineering}} \bibinfo{volume}{5}, \bibinfo{number}{1} (\bibinfo{year}{1998}), \bibinfo{pages}{46--55}.
\newblock
\urldef\tempurl%
\url{https://doi.org/10.1109/99.660313}
\showDOI{\tempurl}


\bibitem[Google(2015)]%
        {gperftools}
\bibfield{author}{\bibinfo{person}{Google}.} \bibinfo{year}{2015}\natexlab{}.
\newblock \bibinfo{title}{Google Performance Tools}.
\newblock
\newblock
\urldef\tempurl%
\url{https://github.com/gperftools/gperftools}
\showURL{%
\tempurl}


\bibitem[Grauer-Gray et~al\mbox{.}(2012)]%
        {polybench}
\bibfield{author}{\bibinfo{person}{Scott Grauer-Gray}, \bibinfo{person}{Lifan Xu}, \bibinfo{person}{Robert Searles}, \bibinfo{person}{Sudhee Ayalasomayajula}, {and} \bibinfo{person}{John Cavazos}.} \bibinfo{year}{2012}\natexlab{}.
\newblock \showarticletitle{{Auto-tuning a high-level language targeted to GPU codes}}. In \bibinfo{booktitle}{\emph{2012 Innovative Parallel Computing (InPar)}}. \bibinfo{pages}{1--10}.
\newblock
\urldef\tempurl%
\url{https://doi.org/10.1109/InPar.2012.6339595}
\showDOI{\tempurl}


\bibitem[Jog et~al\mbox{.}(2015)]%
        {mafia}
\bibfield{author}{\bibinfo{person}{Adwait Jog}, \bibinfo{person}{Onur Kayiran}, \bibinfo{person}{Tuba Kesten}, \bibinfo{person}{Ashutosh Pattnaik}, \bibinfo{person}{Evgeny Bolotin}, \bibinfo{person}{Niladrish Chatterjee}, \bibinfo{person}{Stephen~W. Keckler}, \bibinfo{person}{Mahmut~T. Kandemir}, {and} \bibinfo{person}{Chita~R. Das}.} \bibinfo{year}{2015}\natexlab{}.
\newblock \showarticletitle{Anatomy of GPU Memory System for Multi-Application Execution}. In \bibinfo{booktitle}{\emph{Proceedings of the 2015 International Symposium on Memory Systems}} (Washington DC, DC, USA) \emph{(\bibinfo{series}{MEMSYS '15})}. \bibinfo{publisher}{Association for Computing Machinery}, \bibinfo{address}{New York, NY, USA}, \bibinfo{pages}{223–234}.
\newblock
\showISBNx{9781450336048}
\urldef\tempurl%
\url{https://doi.org/10.1145/2818950.2818979}
\showDOI{\tempurl}


\bibitem[Khairy et~al\mbox{.}(2020)]%
        {accelsim}
\bibfield{author}{\bibinfo{person}{Mahmoud Khairy}, \bibinfo{person}{Zhesheng Shen}, \bibinfo{person}{Tor~M. Aamodt}, {and} \bibinfo{person}{Timothy~G. Rogers}.} \bibinfo{year}{2020}\natexlab{}.
\newblock \showarticletitle{{Accel-Sim: An Extensible Simulation Framework for Validated GPU Modeling}}. In \bibinfo{booktitle}{\emph{2020 ACM/IEEE 47th Annual International Symposium on Computer Architecture (ISCA)}}. \bibinfo{pages}{473--486}.
\newblock
\urldef\tempurl%
\url{https://doi.org/10.1109/ISCA45697.2020.00047}
\showDOI{\tempurl}


\bibitem[Lee and Ro(2013)]%
        {parallelGPUSim1}
\bibfield{author}{\bibinfo{person}{Sangpil Lee} {and} \bibinfo{person}{Won~Woo Ro}.} \bibinfo{year}{2013}\natexlab{}.
\newblock \showarticletitle{Parallel GPU architecture simulation framework exploiting work allocation unit parallelism}. In \bibinfo{booktitle}{\emph{2013 IEEE International Symposium on Performance Analysis of Systems and Software (ISPASS)}}. \bibinfo{pages}{107--117}.
\newblock
\urldef\tempurl%
\url{https://doi.org/10.1109/ISPASS.2013.6557151}
\showDOI{\tempurl}


\bibitem[Lee and Ro(2016)]%
        {parallelGPUSim2}
\bibfield{author}{\bibinfo{person}{Sangpil Lee} {and} \bibinfo{person}{Won~Woo Ro}.} \bibinfo{year}{2016}\natexlab{}.
\newblock \showarticletitle{Parallel GPU Architecture Simulation Framework Exploiting Architectural-Level Parallelism with Timing Error Prediction}.
\newblock \bibinfo{journal}{\emph{IEEE Trans. Comput.}} \bibinfo{volume}{65}, \bibinfo{number}{4} (\bibinfo{year}{2016}), \bibinfo{pages}{1253--1265}.
\newblock
\urldef\tempurl%
\url{https://doi.org/10.1109/TC.2015.2444848}
\showDOI{\tempurl}


\bibitem[Lindholm et~al\mbox{.}(2008)]%
        {teslaHotchips}
\bibfield{author}{\bibinfo{person}{Erik Lindholm}, \bibinfo{person}{John Nickolls}, \bibinfo{person}{Stuart Oberman}, {and} \bibinfo{person}{John Montrym}.} \bibinfo{year}{2008}\natexlab{}.
\newblock \showarticletitle{{NVIDIA Tesla: A Unified Graphics and Computing Architecture}}.
\newblock \bibinfo{journal}{\emph{IEEE Micro}} \bibinfo{volume}{28}, \bibinfo{number}{2} (\bibinfo{year}{2008}), \bibinfo{pages}{39--55}.
\newblock
\urldef\tempurl%
\url{https://doi.org/10.1109/MM.2008.31}
\showDOI{\tempurl}


\bibitem[Malhotra et~al\mbox{.}(2014)]%
        {gputejas}
\bibfield{author}{\bibinfo{person}{Geetika Malhotra}, \bibinfo{person}{Seep Goel}, {and} \bibinfo{person}{Smruti~R. Sarangi}.} \bibinfo{year}{2014}\natexlab{}.
\newblock \showarticletitle{GpuTejas: A parallel simulator for GPU architectures}. In \bibinfo{booktitle}{\emph{2014 21st International Conference on High Performance Computing (HiPC)}}. \bibinfo{pages}{1--10}.
\newblock
\urldef\tempurl%
\url{https://doi.org/10.1109/HiPC.2014.7116897}
\showDOI{\tempurl}


\bibitem[Narang and Diamos(2016)]%
        {deepbenchWeb}
\bibfield{author}{\bibinfo{person}{S. Narang} {and} \bibinfo{person}{G. Diamos}.} \bibinfo{year}{2016}\natexlab{}.
\newblock \bibinfo{title}{{DeepBench: Benchmarking Deep Learning operations on different hardware}}.
\newblock
\newblock
\urldef\tempurl%
\url{https://github.com/baidu-research/DeepBench}
\showURL{%
\tempurl}


\bibitem[NVIDIA(2017)]%
        {voltaPaper}
\bibfield{author}{\bibinfo{person}{NVIDIA}.} \bibinfo{year}{2017}\natexlab{}.
\newblock \bibinfo{booktitle}{\emph{{NVIDIA Tesla V100 GPU architecture the world's most advanced data center GPU}}}.
\newblock \bibinfo{type}{{T}echnical {R}eport}. \bibinfo{institution}{NVIDIA}.
\newblock


\bibitem[NVIDIA(2018a)]%
        {cutlass}
\bibfield{author}{\bibinfo{person}{NVIDIA}.} \bibinfo{year}{2018}\natexlab{a}.
\newblock \bibinfo{title}{{CUTLASS: CUDA Templates for Linear Algebra Subroutines}}.
\newblock
\newblock
\urldef\tempurl%
\url{https://github.com/NVIDIA/cutlass}
\showURL{%
\tempurl}


\bibitem[NVIDIA(2018b)]%
        {turingPaper}
\bibfield{author}{\bibinfo{person}{NVIDIA}.} \bibinfo{year}{2018}\natexlab{b}.
\newblock \bibinfo{booktitle}{\emph{{NVIDIA TURING GPU architecture Graphics Reinvented NVIDIA Turing GPU Architecture}}}.
\newblock \bibinfo{type}{{T}echnical {R}eport}. \bibinfo{institution}{NVIDIA}.
\newblock


\bibitem[NVIDIA(2020)]%
        {amperePaper}
\bibfield{author}{\bibinfo{person}{NVIDIA}.} \bibinfo{year}{2020}\natexlab{}.
\newblock \bibinfo{booktitle}{\emph{{NVIDIA AMPERE GA102 GPU architecture Second-Generation RTX NVIDIA Ampere GA102 GPU Architecture}}}.
\newblock \bibinfo{type}{{T}echnical {R}eport}. \bibinfo{institution}{NVIDIA}.
\newblock


\bibitem[NVIDIA(2022a)]%
        {adaPaper}
\bibfield{author}{\bibinfo{person}{NVIDIA}.} \bibinfo{year}{2022}\natexlab{a}.
\newblock \bibinfo{booktitle}{\emph{{NVIDIA ADA GPU architecture}}}.
\newblock \bibinfo{type}{{T}echnical {R}eport}. \bibinfo{institution}{NVIDIA}.
\newblock


\bibitem[NVIDIA(2022b)]%
        {hopperPaper}
\bibfield{author}{\bibinfo{person}{NVIDIA}.} \bibinfo{year}{2022}\natexlab{b}.
\newblock \bibinfo{booktitle}{\emph{{NVIDIA H100 Tensor Core GPU Architecture}}}.
\newblock \bibinfo{type}{{T}echnical {R}eport}. \bibinfo{institution}{NVIDIA}.
\newblock


\bibitem[Optimisation and (PoP)(2020)]%
        {criticalSectionOpenMP}
\bibfield{author}{\bibinfo{person}{Performance Optimisation} {and} \bibinfo{person}{Productivity (PoP)}.} \bibinfo{year}{2020}\natexlab{}.
\newblock \bibinfo{title}{Patterns of OpenMP critical section}.
\newblock
\newblock
\urldef\tempurl%
\url{https://co-design.pop-coe.eu/patterns/openmp-critical-section.html}
\showURL{%
\tempurl}


\bibitem[Overflow(2009)]%
        {stlThreadSafe}
\bibfield{author}{\bibinfo{person}{Stack Overflow}.} \bibinfo{year}{2009}\natexlab{}.
\newblock \bibinfo{title}{Are C++ STL containers thread-safe?}
\newblock
\newblock
\urldef\tempurl%
\url{https://stackoverflow.com/questions/1362110/is-the-c-stdset-thread-safe}
\showURL{%
\tempurl}


\bibitem[Sun et~al\mbox{.}(2019)]%
        {mgpusim}
\bibfield{author}{\bibinfo{person}{Yifan Sun}, \bibinfo{person}{Trinayan Baruah}, \bibinfo{person}{Saiful~A. Mojumder}, \bibinfo{person}{Shi Dong}, \bibinfo{person}{Xiang Gong}, \bibinfo{person}{Shane Treadway}, \bibinfo{person}{Yuhui Bao}, \bibinfo{person}{Spencer Hance}, \bibinfo{person}{Carter McCardwell}, \bibinfo{person}{Vincent Zhao}, \bibinfo{person}{Harrison Barclay}, \bibinfo{person}{Amir~Kavyan Ziabari}, \bibinfo{person}{Zhongliang Chen}, \bibinfo{person}{Rafael Ubal}, \bibinfo{person}{José~L. Abellán}, \bibinfo{person}{John Kim}, \bibinfo{person}{Ajay Joshi}, {and} \bibinfo{person}{David Kaeli}.} \bibinfo{year}{2019}\natexlab{}.
\newblock \showarticletitle{MGPUSim: Enabling Multi-GPU Performance Modeling and Optimization}. In \bibinfo{booktitle}{\emph{2019 ACM/IEEE 46th Annual International Symposium on Computer Architecture (ISCA)}}. \bibinfo{pages}{197--209}.
\newblock


\bibitem[Ubal et~al\mbox{.}(2012)]%
        {multi2sim}
\bibfield{author}{\bibinfo{person}{Rafael Ubal}, \bibinfo{person}{Byunghyun Jang}, \bibinfo{person}{Perhaad Mistry}, \bibinfo{person}{Dana Schaa}, {and} \bibinfo{person}{David Kaeli}.} \bibinfo{year}{2012}\natexlab{}.
\newblock \showarticletitle{Multi2Sim: A simulation framework for CPU-GPU computing}. In \bibinfo{booktitle}{\emph{2012 21st International Conference on Parallel Architectures and Compilation Techniques (PACT)}}. \bibinfo{pages}{335--344}.
\newblock


\bibitem[Villa et~al\mbox{.}(2021)]%
        {nvas}
\bibfield{author}{\bibinfo{person}{Oreste Villa}, \bibinfo{person}{Daniel Lustig}, \bibinfo{person}{Zi Yan}, \bibinfo{person}{Evgeny Bolotin}, \bibinfo{person}{Yaosheng Fu}, \bibinfo{person}{Niladrish Chatterjee}, \bibinfo{person}{Nan Jiang}, {and} \bibinfo{person}{David Nellans}.} \bibinfo{year}{2021}\natexlab{}.
\newblock \showarticletitle{Need for Speed: Experiences Building a Trustworthy System-Level GPU Simulator}. In \bibinfo{booktitle}{\emph{2021 IEEE International Symposium on High-Performance Computer Architecture (HPCA)}}. \bibinfo{pages}{868--880}.
\newblock
\urldef\tempurl%
\url{https://doi.org/10.1109/HPCA51647.2021.00077}
\showDOI{\tempurl}


\bibitem[Yoo et~al\mbox{.}(2003)]%
        {slurm}
\bibfield{author}{\bibinfo{person}{Andy~B. Yoo}, \bibinfo{person}{Morris~A. Jette}, {and} \bibinfo{person}{Mark Grondona}.} \bibinfo{year}{2003}\natexlab{}.
\newblock \showarticletitle{SLURM: Simple Linux Utility for Resource Management}. In \bibinfo{booktitle}{\emph{Job Scheduling Strategies for Parallel Processing}}, \bibfield{editor}{\bibinfo{person}{Dror Feitelson}, \bibinfo{person}{Larry Rudolph}, {and} \bibinfo{person}{Uwe Schwiegelshohn}} (Eds.). \bibinfo{publisher}{Springer Berlin Heidelberg}, \bibinfo{address}{Berlin, Heidelberg}, \bibinfo{pages}{44--60}.
\newblock
\showISBNx{978-3-540-39727-4}


\end{thebibliography}
